\documentclass[twocolumn,
prl
,superscriptaddress
preprint,
]{revtex4}

\usepackage{graphicx,subfloat,float}
\usepackage{color}
\usepackage{comment}
\usepackage{epstopdf}
\usepackage{notes2bib}
\usepackage{tikz}

\usepackage[caption=false]{subfig}
\usepackage{graphicx,subfloat,float,subfig}
\usepackage{amssymb}
\usepackage{amsmath}
\usepackage{color}
\usepackage{comment}

\usepackage{amsmath,amssymb}

\usepackage{array}

\newcolumntype{+}{!{\vrule width 2pt}}

\newlength\savedwidth

\begin{document}
\author{M. Bauer}
\affiliation{%
Arnold-Sommerfeld-Center for Theoretical Physics and Center for NanoScience, \\
Department of Physics, Ludwig-Maximilians-Universit\"at M\"unchen, D--80333 Munich, Germany}
\author{I. R. Graf}
\affiliation{%
Arnold-Sommerfeld-Center for Theoretical Physics and Center for NanoScience, \\
Department of Physics, Ludwig-Maximilians-Universit\"at M\"unchen, D--80333 Munich, Germany}
\author{V. Ngampruetikorn}
\affiliation{Biological Physics Theory Unit, Okinawa Institute of Science and Technology Graduate University, Onna, Okinawa 904 0495, Japan}
\author{G. J. Stephens}
\affiliation{Biological Physics Theory Unit, Okinawa Institute of Science and Technology Graduate University, Onna, Okinawa 904 0495, Japan}
\affiliation{Department of Physics \& Astronomy, Vrije Universiteit Amsterdam, 1081HV Amsterdam, The Netherlands}
\author{E. Frey}
\affiliation{%
Arnold-Sommerfeld-Center for Theoretical Physics and Center for NanoScience,\\
Department of Physics, Ludwig-Maximilians-Universit\"at M\"unchen, D--80333 Munich, Germany}
\email{frey@lmu.de}

\title{Exploiting ecology in drug pulse sequences in favour of population reduction} 

\begin{abstract}
A deterministic population dynamics model involving birth and death for a two-species system, comprising a wild-type and more resistant species competing via logistic growth, is subjected to two distinct stress environments designed to mimic those that would typically be induced by temporal variation in the concentration of a drug (antibiotic or chemotherapeutic) as it permeates through the population and is progressively degraded.  Different treatment regimes, involving single or periodical doses, are evaluated in terms of the minimal population size (a measure of the extinction probability), and the population composition (a measure of the selection pressure for resistance or tolerance during the treatment).  We show that there exist timescales over which the low-stress regime is as effective as the high-stress regime, due to the competition between the two species.  For multiple periodic treatments, competition can ensure that the minimal population size is attained during the first pulse when the high-stress regime is short, which implies that a single short pulse can be more effective than a more protracted regime.  Our results suggest that when the duration of the high-stress environment is restricted, a treatment with one or multiple shorter pulses can produce better outcomes than a single long treatment.  If ecological competition is to be exploited for treatments, it is crucial to determine these timescales, and estimate for the minimal population threshold that suffices for extinction. These parameters can be quantified by experiment.
\end{abstract}
\maketitle





\section*{Introduction}

Despite recent searches for new classes of antibiotics, based on efficient
screening of uncultured bacteria (or genomes)~\cite{Lok, FischbachWalsh09},
the decline in the rate of development of new types of antibiotic classes since the 1960s, and the concurring increase
in drug-resistant organisms, is disconcerting~\cite{LevyMarshall04, Norrby2005}. 
Growing fears that the world may be re-entering the `prebiotic era'~\cite{Davies2010} has prompted the World Health Organisation
to publish a global action plan in 2015. 

This plan explicitly underlines the importance of optimising current treatment strategies. This encompasses, for example, avoidance of the unwarranted prescription of antibiotics~\cite{Spellberg08}, since their use confers a selective advantage on the resistant variant. Practices include the use of redundant broad-spectrum antibiotics, or longer treatment durations than required to eliminate the infection~\cite{Taubes, Leekha11}.
The optimal duration and intensity of treatments, with antibiotics or drugs in general, are subjects of controversial debates, 
particularly for cases  where one already expects a mutant to be present initially~\cite{Read2011,  LevinRozen}. 
Of course, the ideal treatment should resolve the tension between reducing the advantage enjoyed by the resistant bacteria by preserving the other bacteria in the biome - best achieved by short, mild or specific application of antibiotics - and conclusively eliminating the infection - best attained by long-term application of high doses of antibiotics of assured effect~\cite{Rexconsortium}.
 
In this work, we compare antibiotic treatments, which we refer to as pulses or pulse sequences, with respect to their efficiency in reducing the population size of an infectious two-species population, consisting of one wild-type and one more resistant species.
In our model, treatments with the same concentration profile and treatment length, but different numbers of pulses, are compared with each other, e.g. two shorter pulses compared to one longer one. 
A pulse sequence imposes different patterns of high or low stress on the bacteria, 
mimicking the gradual infiltration of the infection site and the slow degradation of the drug. 
Such environmental changes between high and low stress environments have previously been studied in the context of phenotypically heterogeneous populations~\cite{VeeningSmitsKuipers}.
Much theoretical work has investigated  whether and how phenotypic switching can optimise the long-term fitness
of the species under periodic or stochastic environmental variation~\cite{ThattaiOudenaarden, KlumppPatraPhysBiol, SalatheFeldman, Lachmann, KussellLeibler, KussellReview}. We focus on reducing the size of a population in which the phenotypic switching rates are irrelevant apart from determining the initial composition of the population prior to treatment.

	\subsubsection*{Exploiting competition between species to reduce the evolution of resistance?}

To quantify pulse efficiency, we primarily study the minimal population size $n_{\textrm{min}}$ of our two-species system as a proxy for the extinction probability of the population. Antibiotic stewardship programmes suggest that for some diseases, such as pneumonia, the immune system can clear the residual infection once the bacterial population size is sufficiently reduced~\cite{Mouton11,Hayashi11}.
Thus, the minimal population size may be a more relevant parameter than the exact extinction probability itself. Additionally, the general behaviour of the deterministic system and its observable $n_{\textrm{min}}$ is more robust than the extinction probability in the stochastic model, as in the latter, the precise form of stochastic noise, or the system size, would be important. The total population minimum $n_{\textrm{min}}$ can still serve to gauge the latter, which scales as $\exp(n_{\textrm{min}})$~\cite{vanKampen}.

\begin{figure}[H]
\includegraphics[width=.9\linewidth]{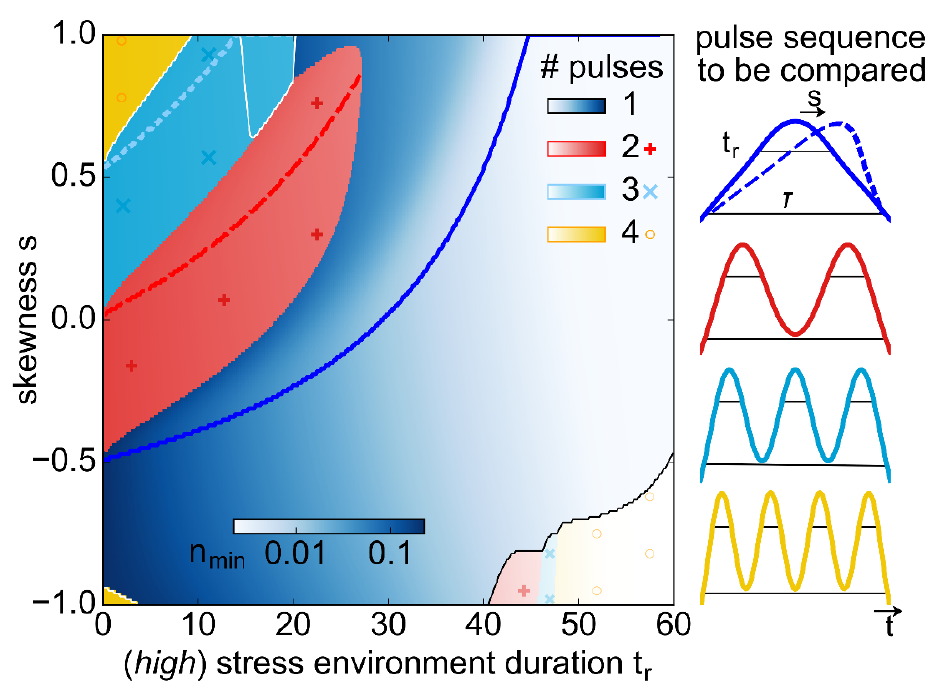}
\caption{\label{fig:allpulse} 
	Comparison of the lowest population minimum $n_{\mathrm{min}}$ for drug treatment sequences of constant overall duration $\tau=60$ providing the same total exposure to high stress $t_r$ and low stress $\tau-t_r$ distributed over different numbers of identically-shaped pulses $N$. Parameters $\tau$ and $t_r$ (x-axis), both given in units of growth rate, together with the skewness $s$, describing how  $t_r$ is positioned within $\tau$, (y-axis), determine a drug pulse (sketched on the right). The colour code signifies the optimal $N$-pulse sequence for a certain choice of parameters $t_r$ and $s$, and the shade indicates the value of the $n_{\mathrm{min}}$. The coloured lines mark the skewnesses that give local minima for fixed $t_r$ corresponding to an optimal onset time for the respective number of pulses. In the non-blue regions a sequence of pulses is more effective than a single pulse, either because onset times are closer to the optimum value for higher number of pulses compared to one pulse (top left corner) or because the minimum $n_\textrm{min}$ is reached in the last pulse of every sequence (bottom right corner; region surrounded by the black line).
}
\end{figure}

Before introducing our model in detail, we jump ahead and summarise the essential result of our work in Fig.~\ref{fig:allpulse}, which we discuss later in more detail.
Fig.~\ref{fig:allpulse} shows the value of the minimal total population size in the configuration space of drug concentration profiles, spanned by the width of the pulse (or duration of its high stress environment) on the x-axis and the form of the pulse on the y-axis, which will be explained later. Practically, these two properties of a pulse - its high stress duration and its form - are likely constrained: a very long duration of the high stress environment or stronger drug might be detrimental for patients due to, for example, a destructive impact on the gut microbiome \cite{Sekirov2008, Shashkova2016}. Similarly, some pulse forms, such as those where the highest possible drug concentration suddenly drops to zero at the pathogen location at the end of the pulse (here denoted by temporal skewness $s=1$), may not be realistic for clinical treatments.  However, since we do not want to make any assumptions on which parts of the configuration space should be accessible, we examine our system for all possible combinations of pulse form and durations of the high stress environments. 

The colour code (symbols) signifies which of the four possible pulse sequences sketched on the right of Fig.~\ref{fig:allpulse} most effectively reduces the population size.
Fig.~\ref{fig:allpulse} clearly shows that in our simple model setup, 
different pulse sequences are favourable in different regions of configuration space. The aim of this work is to outline phenomenologically which pulse sequence yields the lowest minimal population 
for which part of configuration space in Fig.~\ref{fig:allpulse}, and might therefore be most likely to drive the species to extinction. 
The best pulse sequence at any one point of Fig.~\ref{fig:allpulse} tends to be the one that maximally exploits the competition between the more resistant and wild-type species, represented by logistic growth in our model. The simplicity of our approach makes explicit why some references might argue for more moderate treatments involving e.~g.~ shorter or lower drug concentrations, but also what the limitation of models and observables are and hence why such moderate treatments may not work in real setups.
We also examine how the population composition (a measure of how strongly the more resistant species dominates the population) evolves, should such a pulse sequence not lead to achieving extinction. 
Finally, we highlight the need for microbial experiments in such temporally varying drug gradients, in order to evaluate the applicability of simple models to real systems.

\section*{Materials and Methods}

\subsection*{Deterministic model for two-species birth-death process}

The simplest model that can be used to study the effect of the temporal concentration profile on 
a heterogeneous population ($n$) consists of two phenotypically different
species, a susceptible ``wild type'' species ($w$), and a more tolerant or resistant species ($r$). 
Its increased resistance comes at the cost of a reduced fitness in the drug-free
environment, which is reflected in a smaller growth rate. 
As in previous works~\cite{Greulich2012, HermsenHwaPNAS}, we assume that the drug is bacteriostatic, that is, it only affects growth, such that growth of each species ceases as soon as its minimum inhibitory concentration (MIC) is exceeded.

\begin{figure}[h]
\centering
\includegraphics[width=.9\linewidth]{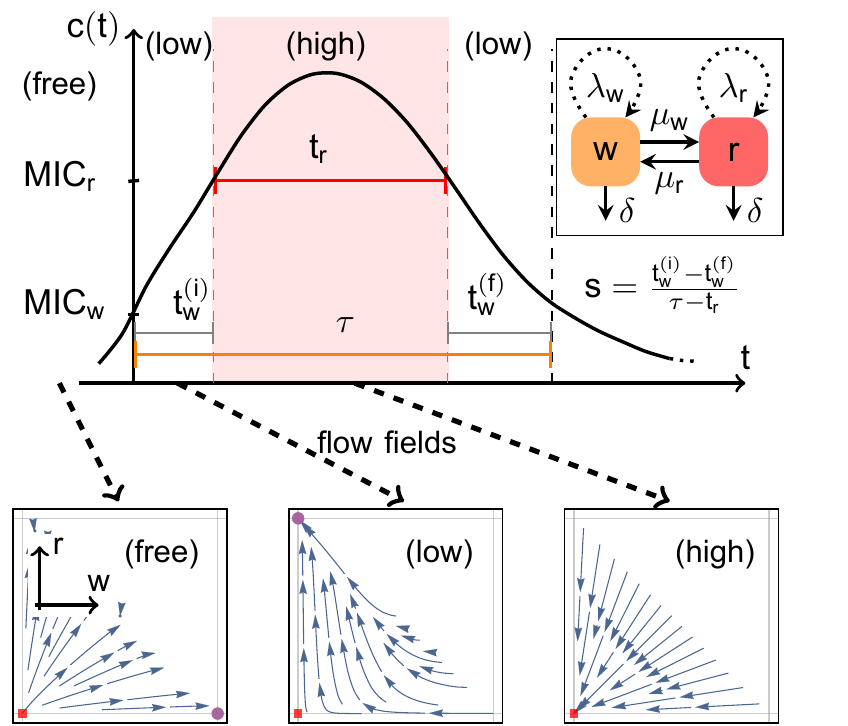}
\caption{
	Top: Inset: Schematic of the two-species model: wild-type $w$ and resistant $r$ grow logistically at rates $\lambda_w$ or $\lambda_r$, decay at rate $\delta$ and switch between states at rates $\mu_w$ or $\mu_r$, respectively. Main figure: Time-dependent antibiotic pulse shape with the three parameters $\tau$, $t_r$, and the skewness $s$ as before. During $t_r$ the antibiotic concentration $c(t)>\mathrm{MIC}_r$ of the more resistant species (\textit{(high)} environment), while during the entire treatment duration $\tau$, $c(t)> \mathrm{MIC}_w$ (\textit{(low)} and \textit{(high)} environment). Initially, the system is in the stress-free environment \textit{(free)}. Bottom:   Dynamical landscapes in population phase space corresponding to these three different environments in antibiotic concentration: 
	\textit{(high)} environment,  with one
	attractive fixed point (red dot) at $n=0$; \textit{(low)} environment, with a saddle point at $n=0$ and an attractive fixed point on the $r$ axis; (\textit{(free)} environment:  with
	unstable fixed point at $n=0$ and stable fixed point close to the $w$ axis, $(w^{*}_{(\textrm{\textit{free}})},r^{*}_{(\textit{free})})$, which we
	use as the initial configuration.
}
\label{fig:model}
\end{figure}

Thus, in this deterministic population dynamics model for the birth-death process, sketched in the inset of Fig.~\ref{fig:model}, the growth rate of each species $\eta \in \{r,w\}$ is given by
$\phi_\eta \left(t, n\left(t\right)\right)= \Theta[\text{MIC}_\eta-c(t)] \lambda_\eta (1-n(t))$,
where $n(t)=w(t)+r(t)$ is the total number of species at time $t$ expressed in terms of a carrying capacity, which does not require specification as it serves merely as a unit for the population size. The Heaviside-Theta step function  $\Theta$  implies that the growth rate is only non-zero when the drug concentration is lower than the MIC of the corresponding species.
The index $\eta \in \{r,w\}$ refers to the type of species (resistant or wild-type), and $\lambda_\eta$ is its growth rate.
The more resistant species has a lower basal growth rate in the drug-free environment, i.e., $\lambda_r=\lambda_w-k:=\lambda-k$, where $k>0$ can be interpreted as a cost that the resistant species incurs for being more resistant.  
The logistic growth assumed in this model introduces competition between the wild-type and the resistant species  for limited space and/or resources,
and places an upper bound on the population size. 
We also include a constant death rate $\delta$ for both species, meaning that
a species decays at rate $\delta$ when $c(t)>\mathrm{MIC}_\eta$: For these higher concentrations, growth of species $\eta$ is inhibited and, since switching is negligible, the species can only die.
All rates and times in this work are given in units of $\lambda$.

The time evolution of the population can be studied in terms of the differential equations 
\begin{align}\label{eq:diff}
	\dot w(t)= \left[ \phi_w \left(t,n\left(t\right)\right)- \delta - \mu_w \right] w(t) + \mu_r r(t) \\
	\dot r(t) = \mu_w w(t) + \left[\phi_r \left(t,n\left(t\right)\right) - \delta -\mu_r\right] r(t) \nonumber
\end{align}

since for sufficiently large populations stochastic fluctuations can be neglected.
The two species are coupled via the competition from logistic growth, as well as via the switching rates $\mu_w$ and $\mu_r$. 
Phenotypically more resistant states can be characterised by a reduced growth rate, or complete growth arrest, often known as tolerance or persistence~\cite{Balabanpersisters, Wakamoto, KussellBalaban} 
(for a recent review, see Ref.~\cite{BraunerBalaban}).  Provided that $\mu_{w,r}\ll \delta$, which is the case for both mutation and phenotypic switching, our choice of $\mu_w=10^{-6}\lambda$ and $\mu_r=0$ does not qualitatively affect the results.

For this entire work, we used exemplary values of $\delta=0.1\lambda$ and $k=0.1\lambda$, where $\lambda \equiv 1$, i.~e.~we used $\lambda$ as the basic unit of time. We investigate several other combinations of costs and death rates, in particular combinations with the same death rate, but a smaller and larger cost, in the Supporting Information. There, we show that our results and general statements are still valid for these cases. We chose the values of $\delta=0.1$ and $k=0.1$ since this combination allowed us to show the complete and most general picture of possible best pulse shapes in Fig.~\ref{fig:allpulse}. A smaller (yet also biologically possible) fitness cost would not have contained all different scenarios. We ask the reader to refer to the Supporting Information for more details.

\subsection*{Antibiotic pulse form determined by skewness and pulse width}

Since in our model the only relevant information about the antibiotic concentration
is whether it is above or below the MIC of the corresponding species, any pulse sequence 
is fully determined by the temporal arrangement of low-stress (\textit{low}) and high-stress (\textit{high}) environments. In these (\textit{low}) and (\textit{high}) environments,  the antibiotic concentration is low, $\textrm{MIC}_{w}<c(t)<\textrm{MIC}_{r}$, or high, $c(t)>\textrm{MIC}_{r}$, respectively (sketched for a single pulse in the top panel of Fig.~\ref{fig:model}). Before the pulse sequence, the system is in
the drug-free environment (\textit{free}), where the concentration of the antibiotic $c(t)$ is less than either MIC, $c(t) <\textrm{MIC}_{w,r} $. We assume that the (\textit{free}) environment appears only before, but not during, a pulse sequence. Thus, the (\textit{free}) environment determines the initial condition of the population, which
we take to be at its fixed point, $(w(t=0),r(t=0)) = (w^{*}_{(\textit{free})},r^{*}_{(\textit{free})})$, shown as the purple dot
near the $w$-axis of the phase space panel (\textit{free})  of Fig.~\ref{fig:model}.

The change in population size and composition in each of these environments is characterised by the flow field in phase space $(w,r)$, shown in the three lower panels of Fig.~\ref{fig:model}. In the (\textit{low}) environment, the population flows towards the more resistant species (high $r$ and low $w$), while in the (\textit{high}) environment, it flows towards the origin, meaning that both species die out exponentially.

Thus, the effect of a single pulse on the
population crucially depends on the times spent by the system in the (\textit{low}) and (\textit{high}) environments. 
A single pulse involves a single (connected) environment of (\textit{high}) antibiotic exposure, with (\textit{low}) environment potentially preceding or succeeding this (\textit{high}) environment. In reality, the duration of these (\textit{low}) environments  will depend on the experimental setup or host. 
A pulse sequence is composed of a succession of identical single pulses.
We refer to the total time of the pulse as $\tau$, and the time during which the system is in the (\textit{high}) environment as $t_r$. The time periods during which the system is in the (\textit{low}) environment (initially) before $t_r$  and (finally) after $t_r$ are denoted by $t_{w}^{(i)}$ and $t_{w}^{(f)}$, respectively. As this would overparameterise the pulse, we combine the latter two time scales into a skewness parameter $s = (t_{w}^{(i)}-t_{w}^{(f)})/(\tau-t_r)$, signifying how $t_r$ is positioned within $\tau$.
Skewness $s=-1$ ($s=1$) thus denotes a pulse which starts (ends) with the (\textit{high}) environment, while skewness $s=0$ denotes a symmetric pulse.

\section*{Results}
	\subsection*{Minimal population}
We compared  pulse sequences of up to $N=4$ pulses (same $t_r$ and $s$) for constant treatment time $\tau$ for all possible skewnesses $s$ and durations $t_r$.
Thus, a single pulse with $\tau=60$ and given $t_r$  and $s$ is compared with a sequence of $N$ identical pulses, each defined by $\tau^{(N)}=60/N$ and $t_{r}^{(N)}=t_r/N$ and $s$. (Further values of $\tau$ are discussed in the Supporting Information.) 
The retention of the same skewness within a sequence is motivated by the fact that we assume that the rate of increase or decrease in concentration is primarily determined by the host system of the bacteria.
In this comparison, the `best' pulse sequence for given $(t_r,s)$ is defined as the one that yields the lowest population minimum $n_\textrm{min}$ and so has the highest likelihood of eliminating the pathogen. In situations where the entire configuration space is accessible, the maximal $t_r$ yields the overall lowest population minimum, independent of skewness $s$. Since practically the maximal duration $t_r$ acceptable for treatments may be limited, it is important to know which pulse sequence is best for each $(t_r,s)$, such that we can provide intuition on any situation and parameter choice that may arise.
The colour (and corresponding symbols) in Fig.~\ref{fig:allpulse} show the best pulse sequences (i.e. the best $N$), and the shade indicates the value of $n_\textrm{min}$ (dark denotes high values).

\begin{figure}[H]
  {\includegraphics[width=.9\linewidth]{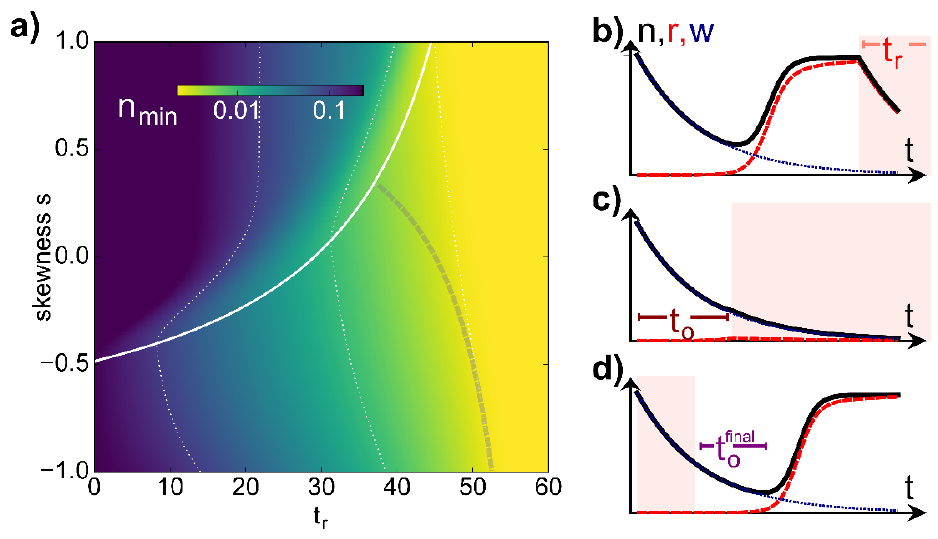}}
	\caption{\label{fig:singlepulse}
	Which value of $s$ gives the lowest population minimum for fixed $t_r$?  a) Lowest population minimum $n_{\mathrm{min}}$ for a single pulse with constant $\tau=60$ for all possible pulse shape parameters $t_r$ and $s$. 
	The optimal skewness $s_o=2t_o/(\tau-t_r)-1$ which gives the smallest  $n_{\mathrm{min}}$ for each $t_r$ is marked in white, while the gray (dashed) line marks the skewness $1-2t_{o}^{\textrm{final}}/(\tau-t_r)$, where $n_{\mathrm{min}}$ has dropped to its smallest value across $t_r$.  The constant contours (dotted lines) serve as guides to the eye. 
	 b-d) Explaining $t_o$ (c, cf. b) and $t_{o}^{\textrm{final}}$ (d), the timescales for which the \textit{(high)} environment is not more effective than the \textit{(low)} environment. In b) $t_{w}^{(i)}$ is longer than $t_o$: the population (black) starts growing during the \textit{(low)} environment, even though the wild-type (blue dotted) decays, as the more resistant species (red) is not affected by the antibiotic. In c), $t_{w}^{(i)}=t_o$ and so the total population keeps decaying. d) If $t_{w}^{(f)}=t_{o}^{\textrm{final}}$ and the pulse ends there, a minimal $n$ is achieved, while for  $t_{w}^{(f)}>t_{o}^{\textrm{final}}$ $n$ grows again.
	}
\end{figure}

We found that a single pulse is most effective over a large range of parameters (blue in Fig.~\ref{fig:allpulse}). In particular, for each duration  in the \textit{(high)} environment $t_r$, the lowest minimum across all skewnesses is obtained by a single pulse (blue line). This means that in practical situations which allow all different pulse skewnesses, a single pulse with a skewness on the blue line would give the lowest minimum. If, however, the possible pulse skewness is limited due to the host setup, a single pulse may not be the best choice. For ease of comparison, Fig.~\ref{fig:singlepulse}~a shows $n_\textrm{min}$ for just a single pulse of constant treatment time $\tau=60$, with the white line marking the lowest minimum (the blue line in Fig.~\ref{fig:allpulse}). In the next paragraph we focus on a single pulse in order to understand which pulse parameters $(s,t_r)$ yield this lowest minimum.

\subsubsection*{Optimal $t_w^{(i)}$ determines lowest possible population minimum}

 We found that this lowest minimum always occurs for pulses with a constant  initial time in the \textit{(low)} environment, $t_w^{(i)}$, which we refer to as $t_o\equiv t_{w,\textrm{optimal}}^{(i)}$. Figs.~\ref{fig:singlepulse}~b-d depict the behaviour of the total population in the (\textit{low}) and  (\textit{high}) environment, with the (\textit{high}) environment marked by the light red background. For a long $t_w^{(i)}$, i.e. a late onset of the (\textit{high}) environment, the dynamics of the total population $n(t)$ (black) and the more resistant species $r(t)$ (red) are shown in Fig.~\ref{fig:singlepulse}~b. Initially $n(t)$ decays exponentially, as the dominating wild-type species dies off. Due to the competition for resources, modelled by logistic growth, $r$ can grow appreciably only once $w$ is sufficiently small. If $t_w^{(i)}$ is long enough, $r$ grows up to the fixed point of the (\textit{low}) environment (see flow in Fig.~\ref{fig:model}). In order to avoid the regrowth of the population, which is then dominated by $r$, the  (\textit{high}) environment should be initiated at the time where $r$ starts dominating and $n$ is minimal (Fig.~\ref{fig:singlepulse}~c).
 This optimal  $t_o \equiv t_{w,\textrm{optimal}}^{(i)}$ depends on the system parameters, and corresponds to $t_o \approx 15$ for our choice of parameters. 
 
 The population can also decay further after the \textit{(high)} environment (Fig.~\ref{fig:singlepulse}~d). Analogously to $t_{w,\textrm{optimal}}^{(i)}$, there exists an optimal $t_{o}^{\textrm{final}} \equiv t_{w,\textrm{optimal}}^{(f)}$, marking the time after which $r$ begins to dominate in the (\textit{low}) environment at the end of the pulse. This then depends on $t_r$ and $t_{w}^{(i)}$. The grey dashed line in Fig.~\ref{fig:singlepulse}~a marks the skewness corresponding to $t_{o}^{\textrm{final}} (t_{w}^{(i)}=0)$ such that $t_{o}^{\mathrm{final}} + t_r = \tau$. For negative $s$, this line indicates where $n_\textrm{min}$ saturates as a function of $t_r$, such that a larger $t_r$ does not yield any drastic changes in $n_\textrm{min}$. 
This means that due to the competitive growth, the \textit{(high)} environment ($t_r$) can always be shorter than \textit{(low)} environment ($\tau$) while still having essentially the same effect as if $t_r=\tau$. A longer duration of \textit{(high)} antibiotic stress is thus not necessarily more effective in reducing the bacterial population, while potentially being disadvantageous for the host.

 For a  survey of the dependence of both $t_o$ and $t_{o}^{\textrm{final}}$ on the system parameters, please refer to the Supporting Information. %

	\subsubsection*{For pulse sequences, the global population minimum can occur during any pulse.}

\begin{figure}[h]
  \includegraphics[width=.9\linewidth]{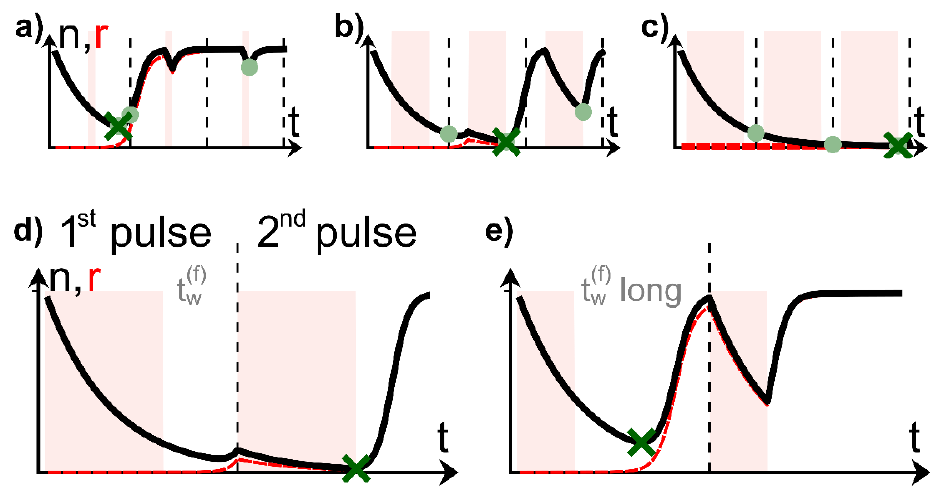}
	\caption{\label{fig:sketch}
	a-c) For a sequence of three pulses, the global population minimum can occur during any of the pulses, depending on the pulse parameters (a: $t_r=6, s=0$, b: $t_r=30,s=0$, c: $t_r=45,s=-0.2$; $\tau=60$ for all). The local minimum (within one pulse) is marked with a light green dot, the global minimum is marked with a green cross. In a, the value of $n$ at the minima (green dots) increases successively, such that global minimum occurs during the first pulse. This is because $\tau - t_r$ is large. For b and c, $\tau -t_r$ decreases, implying that the global minimum shifts into the second and third pulse, respectively. Panels d-e show that for $s=-1$, the minimum is attained in the second pulse (d), unless $t_{w}^{(f)}$ is so long that the population can regrow to $n=r^{*}_{\textrm{\textit{low}}}$ (e).
	}
\end{figure}

When comparing pulse sequences with different $N$ (number of pulses), 
it is important to note that every pulse produces a local population minimum, and the global population minimum can occur during any of these pulses
(see green cross in Fig.~\ref{fig:sketch}~a-c). In the following, we outline heuristically during which pulse the global minimum is likely to occur, and thus which $n_\textrm{min}$ (from which pulse) should be compared to the $n_\textrm{min}$ from the single pulse, or the other sequences.
In practice, predicting which pulse leads to the lowest minimum and hence the highest extinction probability may well be important for treatment: if a later pulse leads to a much larger population reduction than the first, longer treatments with antibiotics are probably beneficial, while if the opposite holds true, the continuation of the treatment may be superfluous or even deleterious.

The population minimum attained during a pulse depends both on the total population size $n$ at the beginning of the pulse, and on the effective duration of population decay. Later pulses in a sequence can in principle have a lower starting population $n$ than the initial pulse, if $w$ has already decayed and $r$ not yet grown substantially (see section Population Composition). 
However, the effective duration during which the population decays shortens from pulse to pulse, simply because $w$ decreases: for example, we already saw that in a single pulse, the population can decay for $t_o + t_r$ or $t_{o}^{\textrm{final}} + t_r$ for $s=1$ and $s=-1$. Both of these time periods of population decay are longer than $t_r$, the effective duration of decay in a pulse late in the sequence, where the population is entirely dominated by $r$. 

Heuristically, the global minimum occurs during later pulses of the sequence for small $\tau-t_r$ (Fig.~\ref{fig:sketch}~c), corresponding
to short periods per pulse spent in the \textit{(low)} environment. This is because, in this case, the population $n$ is small, as $r$ 
cannot have grown drastically, and is additionally depleted during long stretches of $t_r$. Conversely, for small $t_r$ (Fig.~\ref{fig:sketch}~a) the treatment is most efficient in the first pulse, as $r$ progressively takes over the population in the \textit{(low)} environment of each succeeding pulse.
Thus, in the top left corner of Fig.~\ref{fig:allpulse}, we compare the minima from earlier pulses or the first pulse
in each sequence, while in the bottom right corner, it is the later minima in the sequence that need to be compared.

\subsubsection*{For $t_r\ll \tau$ and high $s$, higher $N$ pulse sequences do better, in fact already in the first pulse.}

With this in mind, we are now able to discuss the different features and regions of Fig.~\ref{fig:allpulse}.

First, we focus on the coloured sectors in the \textit{top left corner  of Fig.~\ref{fig:allpulse}}, corresponding to the region of parameter space where the best pulse sequence consists of two or three pulses. We found that in the part of the slice where the global minimum occurs in the first pulse (e.g. the red slice), the minimal population $n_\textrm{min}$ can assume a minimum for each duration of the \textit{(high)} environment $t_r$, similar to the blue line for the single pulse. This minimum arises because the initial \textit{(low)} environment, with duration $t_{w}^{(i)}$, can be exploited to maximise population decay in longer pulse sequences as well. 
In order to know how long this \textit{(low)} environment lasts for the first pulse of a longer sequence, we need to note that the first pulse in a two pulse sequence is only half as long as the full pulse for the single pulse. Thus, 
all relevant time scales need to be rescaled by $N$, i.e. the full pulse duration of a first pulse in a $N$-pulse sequence is $\tau^{(N)}=\tau/N$, the duration of the \textit{(high)} environment is $t_{r}^{(N)} = t_r/N$, and the durations in the initial and final \textit{(low)} environment are $t_{w}^{(i/f),(N)}=t_w^{(i/f)}/N$, where $t_w^{(i/f),(N)}$ denotes the time $t_w^{(i/f)}$ in one pulse of a $N$-pulse sequence.  Then, the skewness at these minima for fixed $t_r$, marked by the red (dashed) and light-blue (dotted) lines in Fig.~\ref{fig:allpulse},
is given by $ s^{(N)}_o=2 N t_o/ (\tau-t_r)-1$ for $N$ pulses. 
As a result of this multiplication by $N$ compared to the optimal skewnesses for a single pulse, the lines marking the lowest population minimum within a region are shifted upwards for higher $N$ sequences~\footnote{We remark on the side that the reason for why the red region does not extend all the way up to $s=1$ is that a single pulse with a long $t_r^{(1)} > t_o +t_r^{(N)}$ will yield a lower $n_\textrm{min}$ just by decreasing $r$ (because the single pulse makes the population decay for at least $t_r$, whereas the initial pulse in the $N$-pulse sequence only decreases the population for $t_o +t_r^{(N)} = t_o + t_r/N$). This also applies to higher $\tau >60$, such as shown in the Supporting Information.}.

In the \textit{bottom right corner of Fig.~\ref{fig:allpulse}}, marked with a thin black line, the last pulse in the sequence yields the lowest minimum. This region is located at negative $s$, indicating that now the time spent in the final \textit{(low)} environment, $t_w^{(f),(N)}$, is important. Indeed, this can be seen clearly by considering a sequence of two pulses with $s=-1$, such as sketched in Fig.~\ref{fig:sketch}~d-e: the lack of an initial \textit{(low)} environment \textit{can} mean that a second pulse is better, because the population could in principle decay for a consistent run of the \textit{(high)} and \textit{(low)} environments of the first pulse, and \textit{(high)} environment of the second pulse, which together corresponds to a time of $t_r^{(N)} +t_w^{(f),(N)} + t_r^{(N)}$  (Fig.~\ref{fig:sketch}~d).
Indeed, for the second minimum not to be lower, the time in the \textit{(low)} environment at the end of a pulse of the $N$-pulse sequence, $t_w^{(f),(N)}$, needs to be long enough for
$r$ to have grown up to $r^{*}_{\textit{(low)}}$ before the onset of the second \textit{(high)} environment of duration $t_r^{(N)}$ (Fig.~\ref{fig:sketch}~e). We note that such a long effective time of population decay is not possible for $s=1$, where the second pulse would begin with a second  \textit{(low)} environment, during which $r$ can regrow.
A more detailed argument, with an estimate for the $t_r$ at which the second pulse provides a lower minimum (i.e. the beginning of the thin black line), is given in the Supporting Information.
For increasing $t_r$, the lowest minimum shifts to even higher $N$, so pulse sequences with increasing $N$ perform better. In addition, the regime where $N>1$ pulse sequences yield lower minima increases along the $s$-axis, until at $t_r= \tau$  all pulse sequences correspond to the same pulse and the minimum is equal across all skewnesses. 

Fig.~\ref{fig:allpulse} also shows \textit{regions marked with a white line}, where the lowest minimum occurs in an intermediate pulse in the sequence. We do not discuss this further here, but note for completeness that had we included longer pulse sequences, these regions would have split up further and shown that longer pulse sequences would give even lower minima in an intermediate pulse. For a better overview of the behaviour of regions where an intermediate pulse does best, we refer to Fig.~A and B in the Supporting Information S1 Text. 
The exact position of this region varies, if higher $N$ pulse sequences are included, for different choices of fitness costs, death rate, and switching rate, and for different $\tau$. Thus, a detailed analysis is not worthwhile, especially as the differences in the numerical values between the lowest minimum of a higher $N$-sequence, and the first minimum of a shorter sequence, are at least two magnitudes lower than the absolute value of $n_\textrm{min}$  there.

\subsection*{Population Composition \label{sec:pc}}

\begin{figure}[h]
  {\includegraphics[width=.9\linewidth]{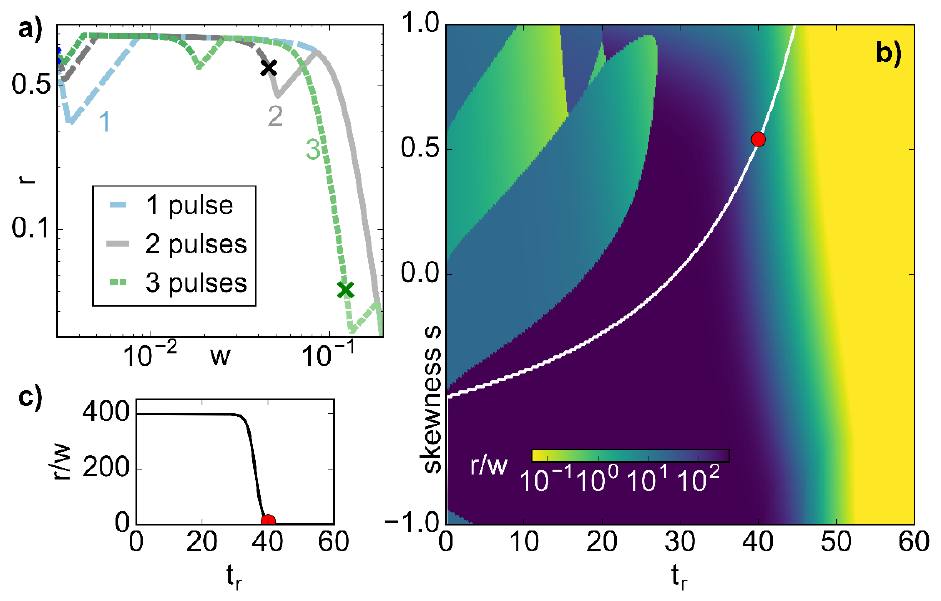}}
	\caption{\label{fig:spirals} 
	Panel a) Phase space $(w,r)$ trajectories for parameters $\tau=60$, $t_r=10$, $s=0.9$ and different numbers of pulses. The trajectories start from the fixed point of the antibiotic-free environment $(w^{*}_{\textit{(free)}},r^{*}_{\textit{(free)}})$, with $r/w$ increasing with every pulse. Here, the sequence with three pulses  gives the lowest minimum [see also Fig.~\ref{fig:allpulse}]. The cross marks the end of the best pulse  within each sequence.  Panel b) Ratio $r/w$ at the end of the 
	best pulse of the best sequence from Fig.~\ref{fig:allpulse} . For fixed $t_r$, $r/w$ is best (smallest) in the regions where the first pulse in a high $N$ pulse sequence yields the lowest $n_\textrm{min}$ (top left corner, also in Fig.~\ref{fig:allpulse}). Similarly to Fig.~\ref{fig:singlepulse}, the white line marks the skewness corresponding to the lowest minimum for each $t_r$, with the absolute value of the ratio dropping drastically at the red dot, shown in c). Panel c) Dependence of $r/w$  on $t_r$ along the white line in b). The ratio is approximately constant for $t_r<30$ and then suddenly drops to the same value that it would also show at $t_r = \tau$ 
}
\end{figure}

In the previous section, we learnt which pulse sequences  
 yield the maximal relative reduction in population size for which regions in $(t_r,s)$-space. This minimal population
 $n_\textrm{min}$ served as a proxy for gauging when extinction would most likely occur in a setting where an immune response can destroy
 the population when it is already small. Now, we would like to address a complementary question: in the event
 that extinction does not occur, whether because $n_\textrm{min}$ was too high, or the population was small for too short, what is the effect of such a `failed' pulse on the bacterial population?
We already saw that the composition of the
population shifts more towards $r$ with each pulse.
In terms of real treatments, it might often be better not to pursue treatments which, if unsuccessful, entail a high risk of creating a fully resistant population. 
In order to evaluate the pulsed treatments associated with the most effective population reductions based on Fig.~\ref{fig:allpulse},
we now focus on the population composition, quantified by the ratio of resistant to wild-type species, $r/w$,
at the end of the best pulse within the best sequence. 
Evaluating $r/w$ at the end of the pulse that yields the global minimum is motivated by the fact that the treatment can be stopped after, but not during, an individual pulse. Fig.~\ref{fig:spirals}~b shows the dependence of $r/w$ on the pulse configuration, which
can be best understood by first considering how the population evolves in $(w,r)$ phase space during the different pulse sequences. 

In Fig.~\ref{fig:spirals}~a, we show trajectories for three pulse sequences, consisting of a single, two, or three pulses respectively, with $\tau=60$ and $t_r=10$.
The qualitative behaviour of the phase space trajectory is independent of skewness (in Fig.~\ref{fig:spirals}~a, $s=0.9$, the corresponding trajectories for $s=-0.5$,  $s=0.2$ and $s=0.5$ can be found in Fig.~D in S1 Text).
The colour of the trajectory darkens progressively with every pulse in the sequence. 
The trajectory starts at the \textit{(free)} fixed point close to the $w$-axis beyond the limits of  Fig.~\ref{fig:spirals}~a, and evolves towards the $r$-axis.
Within each sequence, $r/w$ steadily increases from pulse to pulse, as $r$ progressively takes over the population during the $\textit{(low)}$ regimes. Thus, in the top left corner of Fig.~\ref{fig:spirals}~b, where the first
pulse of the sequence yields the lowest minimum, $r/w$  is comparatively
smaller (lighter shading). Indeed,  the higher the $N$ of the best sequence, the lower the ratio in Fig.~\ref{fig:spirals}~b, provided the global minimum is reached in the first pulse (such as in the red region in Fig.~\ref{fig:allpulse}). The region marked with the white line in Fig.~\ref{fig:allpulse}, where intermediate pulses (and not the first pulse) in the sequence yielded the lowest global minimum, also shows up clearly as darker in Fig.~\ref{fig:spirals}~b. Here, $r$ has grown more than for a single pulse, as more pulses were applied before the population minimum was reached.

Thus, in our model, when both population reduction and composition are considered, pulse sequences where the minimum is attained in the first pulse 
are generally more effective than a single long pulse: maintaining the \textit{(low)} regime in the first pulse for around $t_o$ keeps $r/w$ as well as $n_\textrm{min}$ small. This argument suggests that treating with this first pulse only achieves the best result, and additionally comes with a shorter total treatment duration $\tau$ and a shorter $t_r$.
We would like to note that even if the population does not die out during this short treatment, multiple pulses of this form could be added in order to give the immune system more opportunities to eliminate the infection. These additional pulses would not drastically change $r/w$ compared to the composition obtained after the single long pulse of $\tau=60$. This can be seen also in Fig.~\ref{fig:spirals}~a, where for all pulse sequences the population composition is similar at the end of the entire sequence.

\subsubsection*{Single pulse ratio for optimal skewness drastically drops to a low value at  $t_r<\tau$}

Finally, we wish to make an observation concerning the population composition following a single pulse, which dominates
Fig.~\ref{fig:spirals}~b. Focusing on the dependence of the minimal $n_\textrm{min}$ on $t_r$ (white lines in Fig.~\ref{fig:singlepulse} and Fig.~\ref{fig:spirals}~b, blue line in  Fig.~\ref{fig:allpulse}), we see that the population composition along this line drops drastically at a certain $t_r< \tau$, as shown in Fig.~\ref{fig:spirals}~c (marked with a red dot). 
This means that also in terms of keeping the resistant species at bay, it is not necessary to impose the \textit{(high)} environment for the entire duration of the pulse sequence.

\subsubsection*{Caveat: Fixing total  antibiotic load, rather than $\tau$, might reduce dependence on pulse form}

So far, we have considered a constant drug expose time $\tau$, corresponding to the entire treatment duration,
which is also the quantity minimised in antibiotic stewardship programs.
Alternatively, it could be important for medical applications to compare different pulses which keep the total load of antibiotic applied constant.
Since our model does not incorporate any information relating to the antibiotic concentration, this question cannot be adequately addressed.
Nevertheless, to a first approximation, $\tau + \alpha t_r$ might be a meaningful proxy for this total load, where $\alpha$ quantifies how $\mathrm{MIC}_w$ relates to $\mathrm{MIC}_r$: 
$\alpha=\frac{\mathrm{MIC}_r - \mathrm{MIC}_w}{\mathrm{MIC}_w}.$

In order to gauge the effect of applying a constant drug load, we fixed $\alpha = 1$ for simplicity, and show both population minima and composition for constant $t_{\mathrm{sum}} = \tau + t_r$ for a single pulse in the Supporting Information. Suffice it to say here that both quantities depend primarily on $t_r$, and only weakly on the skewness of the pulse. This is due to the fact that for $s=1$, a constant $t_\textrm{sum}$ exploits the optimal $t_w^{(i)}$, while for $s=-1$, the optimal $t_w^{(f)}$ is used. Since these quantities are not very different for our choice of parameters, constant $t_\textrm{sum}$ effectively means that even though the skewness is different, the pulse will involve approximately the same $\tau$ and $t_r$ for $s=1$ and $s=-1$. A definite conclusion as to the effect of the form of the pulse on population reduction and composition for fixed antibiotic load is not possible within the scope of this work, and would in any case require a different model.

\section*{Discussion}

\subsubsection*{Experimental Relevance}

Experiments with microbes can help investigate minimal antibiotic dosages and treatment times in a well-controlled test tube setup, where  
the impact of certain treatments on the microbial species itself can be studied without interfering effects, 
for example from the immune system.
Such microbial experiments have, for example, helped suggest drug combinations or treatment regimens which could 
retard the development of antibiotic resistance~\cite{Bollenbach15,ChaitKishony, MichelMoellerKishony, EvgrafovSommer}.
Increasingly, these experiments try to incorporate practically important aspects of heterogeneities in the environment~\cite{KesslerAustinLevine}, 
such as drug concentration gradients.
These gradients can enhance the development of bacterial resistance relative to spatially homogeneous systems~\cite{HermsenHwaPNAS, Greulich2012, Zhang, WuAustin}, as the more resistant species can successfully compete with a faster growing, but more susceptible wild-type species.
Not enhancing the selective advantage of the more resistant species, in the context of temporal heterogeneities in drug levels, including the duration, frequency and even the concentration
profile during a single antibiotic pulse, as studied also in this work, is also important in real treatments~\cite{Taubes,Fishman06}, and is thus within the limits of current experiments.

Our model makes two drastic simplifications compared to real microbial species. First, we study only two species, instead of a series of possible phenotypically or genotypically different species. 
Typically, the evolutionary pathway that leads to a fully resistant species involves a variety of
intermediate mutants, even when the mutational paths are constrained~\cite{Weinreich06}.
Since the fitness benefit diminishes with each successive mutation in a series~\cite{Chou11, Khan11}, 
we assumed that the strongest effect is conferred
by the first mutation, and neglected all higher order mutants.
For phenotypic switches, it is reasonable 
to consider only two species, corresponding to, for example, the expression or repression of a protein~\cite{Gerland2009, LinKussell}. Thus, our model should be applicable to experimental systems, while 
in real patients, different types of tolerant or persister cells might be involved~\cite{FridmanBalaban}, or even interact~\cite{Franspaper}.

The second simplification concerns how these two species are affected by the antibiotic.
In our model, we assume that the antibiotic is bacteriostatic, i.e. only affects the growth of the species~\cite{Greulich2012, HermsenHwaPNAS}.
We also assume that the growth rate of each species falls abruptly to zero when the antibiotic concentration is higher than 
their respective minimum inhibitory concentration (MIC) (see e.g.~\cite{Drusano}). 
The experimental situation is more complex: 
 cessation of growth is not instantaneous, the space occupied by a dead
cell may not immediately become available~\cite{LinKussell},
and the general use of the MIC as an indicator for slow growth is questionable~\cite{ArtemovaGore}.
However, an abrupt change in growth rate at MIC has been verified experimentally for E. coli and chloramphenicol.~\cite{HermsenHwaPNAS}. Additionally, our analysis is based on large numbers rather than extinction events which would be model specific; thus, small changes in the model (such as reduced but non-zero growth rate) should still give qualitatively similar results.

Evaluating the effect of different pulse sequences should be possible within a microfluidics setup, where, for example, periodically fluctuating environments have already been investigated for E. coli and tetracycline~\cite{LinKussell}. We expect that one should be able to observe that the \textit{(low)} environment of drug concentration can be exploited in order to increase extinction probabilty for a \textit{(high)} environment that is present for as short as possible, with the treatment time being constant. How long this duration of the  \textit{(low)} environment is for best exploitation would be sensitive to the growth rate of the more resistant species, which for tetracycline could be generated using a specific promoter, namely the agn43 promotor~\cite{LinKussell,Hasman200, Lim2007}. Just as shown in Fig. \ref{fig:allpulse} , we expect higher $N$ pulse sequences to do better when this duration is optimal for them, but not for the longer pulse. In addition, further study of E. coli in combination with other antibiotics and more resistant strains should also show this, in addition to being more realistic than our simple model.

\section*{Summary and Outlook}

We studied the effect of a temporal pattern of  antibiotic exposure, modelled as alternating periods of high and low stress, on a two-species bacterial population. Our results imply that, in fighting bacterial infections, one can make use of the competition between more and less resistant species. We showed that it is not necessary to impose the high stress regime for the entire duration of the treatment in order to obtain the best possible population depletion and composition. In the context of real infections, where a slow increase or decay of the drug concentration in the environment might be inevitable, it can be reassuring to know that the transient \textit{(low)} stress regime need not be deleterious, and can indeed be exploited. However, if the low stress environment is maintained for too long, with this timescale depending on the fitness of the more resistant strain, the latter will start to grow and eventually dominate.

If the duration of the high stress environment must be minimised (for example, due to its negative impact on other species in the biome), our model suggests that it is beneficial to split the treatment into several individual pulses, such that the total time spent in the \textit{(low)} environment is equally distributed among the pulses (and hence shorter per pulse in the sequence than during a single application). In addition, multiple short treatments tend to be more effective when the absolute time spent in the high stress environment needs to be short. For very long treatment times, on the other hand, a single long application is more efficient.
We expect these main results to be robust against a variety of practically relevant changes in the model, such as a break between treatment times, or a decreasing cost in the growth rate for the more resistant species.

Current medical research argues that more moderate treatments may be beneficial, if mutants are already present in a host. The conventional view has tended to assume that long and aggressive treatments are best at eradicating the infection and reducing the probability of evolving resistance~\cite{Read2011}. In order to address this problem of infections in a host, mathematical and computational models for immunocompetent host have been introduced~\cite{Ankomah2014}, or it has been suggested that an absolute population threshold can be used as a measure for extinction~\cite{KorolevXavierGore}.
Following the latter suggestion, we have tried in this work to gauge the effect of the immune system by understanding how the population can be reduced, but not driven to absolute extinction. Within the limits of our model, our results might support the more moderate approach. Other research focusing on ecological competition between species of different levels of resistance~\cite{DayRead, HansenRead17} gave similar insights: on the population level, long and aggressive treatments reduce the probability of generating mutants, while more moderate treatments can exploit the inter-species competition~\cite{DayRead}. From a physical perspective, it would be interesting to see how this more moderate approach can be reconciled with the physics of small numbers: when extinction is a rare event, the more traditional view supporting aggressive treatments might well be favoured. Either way, in the light of this current debate~\cite{Kouyos2014}, it is important to determine the relevant timescales of growth of the different species in presence or absence of the drug~\cite{Troy2015}, and how the species are coupled, in order to make models more realistic.

In order to obtain estimates of such timescales, it is important to use the knowledge obtained from well-controlled microbial model experiments on bacterial populations both in isolated and host environments (such as, for example, \textit{C. elegans} culture~\cite{Ewbank2011, Moy2006}). These experiments provide an ideal set-up for addressing these bigger questions, as they permit both small and large numbers of species to be studied and the environmental conditions can be precisely monitored. 

\section*{Acknowledgments}
This work was supported by a Marie Sk\l{}odowska-Curie actions Horizon 2020 Fellowship (MB), a DFG Fellowship through the Graduate School of Quantitative Biosciences Munich QBM (IRG), the Okinawa Institute of Science and Technology Graduate University (VN and GS), the Vrije Universiteit Amsterdam (GS), the Deutsche Forschungsgemeinschaft, Priority Programme 1617 “Phenotypic heterogeneity and sociobiology of bacterial populations", grant FR 850/11-1 and the German Excellence Initiative via the programme “Nanosystems Initiative Munich” (EF).\\








%











\section*{Exploiting ecology in drug pulse sequences in favour of population reduction: Supporting information} 




\subsection*{1. Fig. 1 of the main text for different $\tau$}
\begin{figure}[H]
  {\includegraphics[width = 0.95\linewidth]{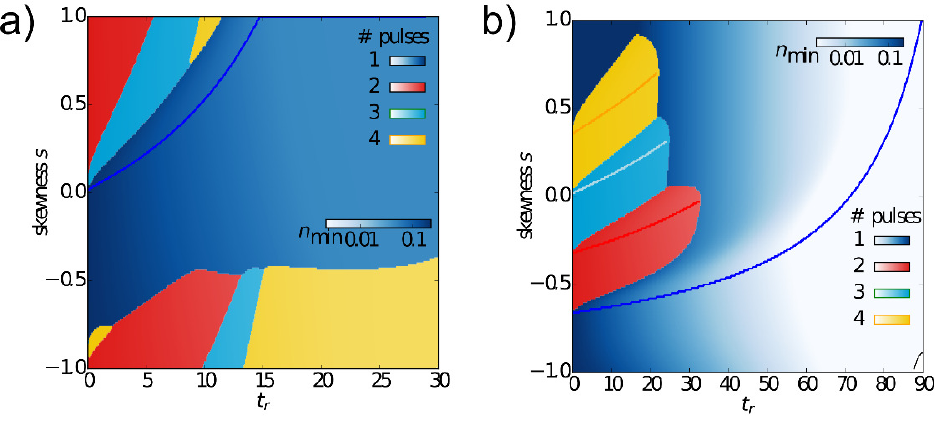}}
	\caption{\label{fig:fig1}
The main figure of the paper, Fig.~1, for different $\tau$: $\tau=30$ for the left (a) and $\tau=90$ for the right panel (b), respectively. We see that the regions where the pulse sequences are better shift consistently: the region where the higher pulses of the sequences yield the lowest global minima (bottom right) shifts towards the left for increasing $\tau$, while the region where the first pulse of the sequence is better than the single pulse (top left) moves inwards and out. In general, for larger $\tau$, the single pulse does better for a larger area in $(s,t_r)$ space.
	}
\end{figure}

In order to gauge how different $N$-pulse sequences compare for different $\tau$, and whether Fig. 1 in the main part of the text changes much, we show the corresponding figures for $\tau=30$ and $\tau=90$ in Fig. \ref{fig:fig1} a and b in S1 Text, respectively.
We see that for smaller $\tau$, the region where the higher $N$-sequences get better increases. This occurs especially for negative $s$, where the region where the final pulses of the higher $N$-sequences are best (bottom right corner, marked with black line in Fig. 1 of the main text) extends towards much smaller $t_r$, and merges with the region of the bottom left, an intermediate pulse in the higher $N$-pulse sequence is best. The extension towards smaller $t_r$ of the region where the last pulse is best can be understood by an argument similar to the main text discussed around Fig. 4 d-e, where one gauges what $t_{w}^{(f)}$ is required for the first pulse to be better; since $\tau$ is smaller, the $t_r$ where $t_r= \tau - t_{w}^{(f)}\equiv \tau-t_o^{\mathrm{final}}$ is smaller too.
For larger $\tau$, the opposite is the case for the bottom right corner, for exactly the same reason. The regions where the first pulse of longer sequences are best moves down towards more negative $s$, but also shrinks in $t_r$ compared to $\tau=60$, as for $\tau=90$ the $t_{r}^{(1)}$ of the single pulse is here already much longer than $t_o + t_{r}^{(N)}$ (see the main text). 
However, the regions do not change qualitatively. 
This might of course be different if one considers $\tau<t_o$ or $\tau<t_{o}^{\mathrm{final}}$, but this would go beyond the scope of this work.

\subsection*{2. Fig. 1 of the main text for different cost for the resistant species, switching rates and death rates}
\begin{figure}
	{\includegraphics[width = 0.95\linewidth]{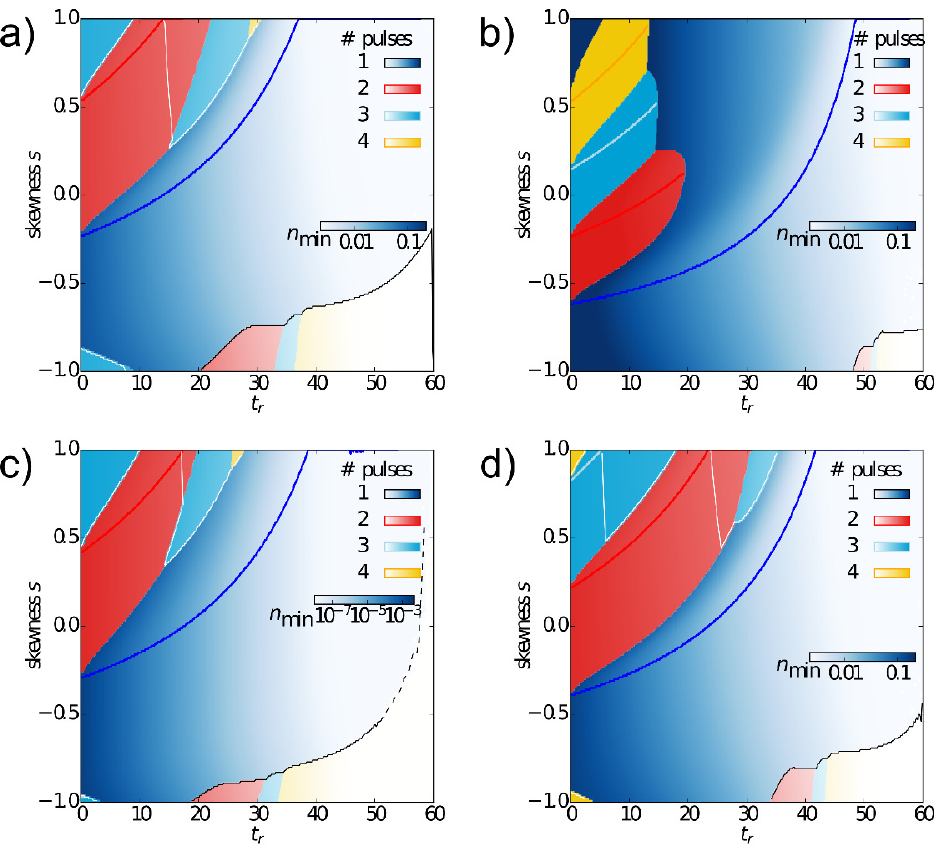}}
		\caption{\label{fig:additional}
	The main figure of this work, Fig.~1, for different rates and costs for mutation: a) $k=0.4, \delta=0.1$; b) $k=0.01, \delta=0.1$; c) $k=0.4, \delta=0.3$; (so far for all $\mu_w=10^{-6}, \mu_r=0$, like in the main text) d) $\mu_w=10^{-7},\mu_r=10^{-4}$ and $\delta =0.1, k=0.1$ like in the main text. The general features are independent of the system parameters. For the largest part of skewnesses and time in the \textit{high} environment, including the region where the skewness is such that initial time in the \textit{low} environment matches the optimal initial time spent in the \textit{low} environment, $t_o$, for the corresponding parameters (thick blue lines), the single pulse is best. The first pulse of a higher order pulse sequence is best when the initial time in the \textit{low} environment of this first pulse matches $t_o$, and in areas around it (top left corner). The last pulse of such higher sequences is best for large $t_r$ and negative skewnesses (marked by a black line). The dominant differences between the different figures are the exact boundaries of these regions, and the regions where intermediate pulses of the longer pulse sequences yield the lowest minimum (white lines). Due to the high decrease in population for the high death rate of $\delta=0.4$, we could not precisely resolve the exact position of the black boundary and thus marked it with dashes. \label{fig:fig1cost}}
\end{figure}
In the main text, we chose a fitness cost of $k=0.1$ for the growth rate of the more resistant species, since this parameter gave a good, yet not too confusing overview over where which pulse sequence yielded the best results. In Fig. \ref{fig:fig1cost}, we show the analogous figures to Fig. 1 in the main text for two different costs $k=0.4$ and $k=0.01$ (top panels, (a) and (b)). These different costs change the values for the optimal initial and final times spent in the \textit{low} environment, $t_o$, and $t_{o}^{\textrm{final}}$ (see section 3 of the supporting information).

The general features of Fig. 1 remain the same: the single pulse is best for the largest part of the phase diagram and is lowest at the thick blue line, where the skewness is such that the initial time in the \textit{low} environment matches the $t_o$ for this parameter combination. The first pulse of a longer pulse sequence is best in the top left corner, around the corresponding thick line where the initial time in the \textit{low} environment for this first pulse matches $t_o$. The region marked with the black curve corresponds to the region where the last pulse of these higher order pulse sequences is best. Compared to Fig. 1 in the main text, the increased cost of $k=0.4$ (Fig. \ref{fig:fig1cost}a in S1 Text) leads to an increased importance of this last region. In addition, the regions where intermediate pulses of the longer pulse sequences are best (marked with white lines), are also larger than in Fig. 1 of the main text, but occur at exactly the predicted positions (in the bottom right corner, and between the regions where first pulses of longer sequences yield lowest minima, in the top left). 

For the smaller cost of $k=0.01$ (Fig. \ref{fig:fig1cost}b) compared to $k=0.1$ in Fig. 1 in the main text, the behaviour is analogous to what we have described so far: the regions encircled in white disappear, and region where the last pulse of the longer sequences yield lower minima (marked with the black line) becomes smaller. In addition, similarly to what happens for longer $\tau$ (see Fig. F in S1 Text), the importance of the region where the single pulse yields best results increases. This is due to the fact that the optimal time spent initially in the \textit{low} environment $t_o$ is shorter for shorter costs, and that thus $\tau/t_o$ increases. Thus means that the $t_r$ of the single pulse is longer, and can reduce the population for longer, than $t_o + t_r/N$ for the $N^{th}$ pulse sequence.

In addition, the changed costs influence the absolute value of the minimal population at the best pulse. The minimal population then is smaller for higher fitness costs, as the population decays more strongly due to the slower growth of the more resistant species.

Finally, we also investigate a combination of increased death rate and cost (Fig. \ref{fig:fig1cost}c, to be compared with the Fig. \ref{fig:fig1cost}a above for the same cost), as well as different switching rates. Again, the features of the figures are exactly the same as in Fig. 1 of the main text. The increased death rate meant that the population reduction was higher, and so for high $t_r$ the different pulse sequences gave very small and effectively indistinguishable lowest minima. Thus, we could not resolve the region where the last pulse of the longer sequences was best precisely, and marked it with a dashed black line in these cases. 

Changing the switching rates (Fig. \ref{fig:fig1cost}d) also leads to only very small changes of general features compared to Fig. 1 in the main text. We note the increase in the areas marked with the white lines, where intermediate pulses of higher pulse sequences do best, but again at positions where they would have occurred for Fig. 1. Thus, we conclude that the main features of, as well as the relevant explanations for Fig. 1 in the main text, are robust against parameter variations.

\subsection*{3. Optimal $t_{w,\textrm{optimal}}^{(i)}=t_o$ and $t_{w,\textrm{optimal}}^{(f)}=t_{o}^{\mathrm{final}}$ for different parameters}

\begin{figure}[H]
  {\includegraphics[width = 0.95\linewidth]{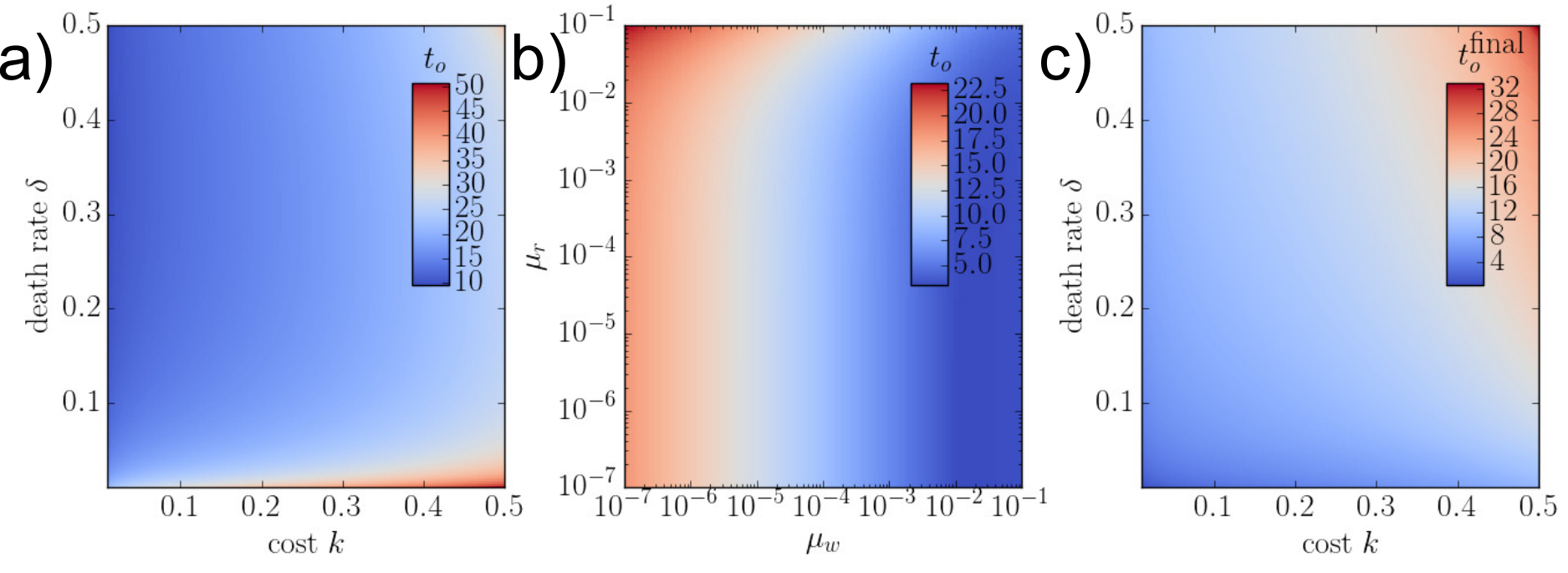}}
	\caption{\label{fig:onset}
	Left panel (a): $t_o$ as a function of death rate $\delta$ and cost $k$ in the growth rate for the more resistant species. For small $\delta$ and large $k$,  $t_0$ can take values of up to $\approx 40/\lambda$, and is thus an important parameter. Middle panel (b): $t_o$ as a function of the switching rates $\mu_r$ and $\mu_w$, which determine only the initial condition given by the attractive fixed point of the \textit{(free)} regime. There is hardly any dependence on $\mu_r$ for $\mu_r\ll \delta$, while $t_o$ decays with increasing $\mu_w$ which implies more resistant agents in the system initially; Right panel (c): Limit of $t_{o}^{\mathrm{final}}(t_w^{(i)}=0)$ for large $t_r$ as a function of death rate $\delta$ and cost $k$. This saturating value of $t_o^{\mathrm{final}}$ increases considerably towards large $\delta$ and large $k$ and is always smaller or equal than $t_0$.}
\end{figure}

In Fig. \ref{fig:onset} we show how the optimal initial and final times in the \textit{low} environment can change with respect to system parameters.

Panel (a) of Fig. \ref{fig:onset} shows the variation of $t_o$ as a function of death rate $\delta$ and cost $k$ in the growth rate of the resistant species, $k=\lambda -\lambda_r$. Obviously, $t_o$ increases when this cost increases, as the resistant species will take longer to grow. For fixed cost, there is a minimal $t_o$ at a certain death rate, with $t_o$ increasing towards both small and large death rates. This is due to the fact that for small growth rate costs for the resistant species, the death rate dominates for the wild type which decays faster, while for large growth rate costs, the resistant species' growth is also drastically limited by the high death rate. This behaviour is most pronounced for large cost $k=0.5$, where $t_o$ diverges for $\delta =0.5$, since then $r$ cannot grow any more at all. In general,  we see that $t_o$ can acquire high high values of around $t_o\approx 50$. This means that the optimal $t_w^{(i)}$ can have a very large effect, and explain why sometimes more moderate treatments are functional, even though they come with the risk of evolving resistance in principle.

In Fig. \ref{fig:onset} b we show how $t_o$ depends on the switching rates $\mu_r$ and $\mu_w$. Even if those rates do not change the actual dynamics very much, they have a big influence on $t_o$, as they crucially influence the initial condition, that is the attractive fixed point of the \textit{(free)} environment. For larger $\mu_w$ the population starts out with a relatively higher amount of resistant species and less wild-type so that the resistant species can take over the population faster, thus decreasing $t_o$. We note that $t_o$ decreases linearly with $\mu_w$ for $\mu_w$ up to $\mu_w\approx 10^{−2}$. The dependence on $\mu_r$ is less pronounced for small $\mu_w$ and $\mu_r$ as to lowest order in $\mu_{w,r}$ the fixed point of the free-system only depends on $\mu_w$.

Concerning $t_{o}^{\mathrm{final}}$, we note that it depends on the preceding $t_w^{(i)}$ and $t_r$. It will be maximal when $t_w^{(i)}=0$, as then $r$ has not yet grown while $w$ has been depleted, and thus if $t_w^{(i)}=0$ and $t_r=0$, $t_{o}^{\mathrm{final}}=t_o$. For $t_w^{(i)}=0$, $t_{o}^{\mathrm{final}}$ decays from this $t_o$ as a function of $t_r$, and saturates at a certain value for high $t_r$, which again depends on the system parameters. In Fig. \ref{fig:onset} c we show the value $t_{o}^{\mathrm{final}}$ saturates to for high $t_r$. We do not show its dependence on $t_w^{(i)}$, as then we would have to additionally satisfy the constraint that $t_w^{(i)} + t_r + t_{o}^{\mathrm{final}} = \tau$, which would distract from the value of $t_{o}^{\mathrm{final}}$. As such, $t_{o}^{\mathrm{final}}$ can take any value between Fig. \ref{fig:onset} a ($t_r=0$) and Fig. \ref{fig:onset} c (large $t_r$).

\subsection*{4. Comparing minima within a pulse sequence}

There are two competing concepts that determine during which pulse the population minimum could occur: 
First, the time during which the population effectively decays exponentially, which can be larger than $t_r$ for pulses early in the sequence where $w$ is still present, and will eventually be $t_r$ for pulses late in the sequence; and second,
the initial population, from which the decay leading to the lowest minimum starts. The latter might be lowest in
intermediate pulses, where $w$ has decayed, but $r$ has not yet taken over the population.

We want to discuss at which $t_r$ a second pulse can yield a lower minimum than a first pulse for $s=1$ and $s=-1$.

For $s=1$, we have seen that the maximal period during which the population decays is $t_w^{(i)}+t_r$ for $t_w^{(i)} \le t_o$ [in principle if $t_w^{(i)}+t_r$ is small, there could be another contribution from $t_w^{(i), 2nd pulse}$, that is a contribution like $t_w^{(f)}$ only at the beginning of the 2nd pulse], which is maximal when $t_w^{(i)}=t_o$. Since pulses are periodic, and since we consider $\tau<126$ 
(which effectively is a requirement on $t_r$ being such that the population will be lower than the minimum of the \textit{(low)} regime (obtained after $t_o$) after the second $t_w^{(i)}$ regime), $n$ will grow up to $r^{low}$ in the following \textit{(low)} regime of the second pulse; there, the population can only decay during $t_r$, and thus the second pulse will have a higher minimum for $t_w^{(i)} = t_o$.
More precisely, for pulses with $t_w^{(i)} > t_o$, the period of effective population decay in the first pulse shortens, and will correspond to $\textrm{max}(t_o,t_r)$ for pulses with $t_w^{(i)} >t_o + t_{regrowth}$, where $t_{regrowth}$ in our case corresponds to approximately $10$. In the \textit{(low)} regime of the second pulse, the population will definitely increase if $t_w^{(i)} > t_o$, meaning that in the second pulse, the effective decay period will only be $t_r$. The decay period of the second pulse is thus either equally long as during the first pulse ($t_r$), or shorter when $t_r<t_o$, with the starting population of decay being definitely at  $r^{low}$. In the latter case, the second pulse will definitely have a higher minimum; in the first, where $t_w^{(i)} > t_o + t_{regrowth}$ and $t_r>t_o$ and the decay periods of both pulses are effectively the same, the second pulse will have an identical minimum to the first pulse. (Also, this latter case only applies for sequences with $\tau_(N) >2*t_o+t_{regrowth} \approx 35$). Thus, the second pulse has a higher or equal minimum whenever $t_w^{(i)} > t_o$.

For $s=-1$, we already noted one can have a very low minimum in the second pulse if the population decays consistently for
$t_r^{(N)} +t_w^{(f),(N)} + t_r^{(N)}$. 
However, the second pulse will actually yield a lower minimum even for larger $t_w^{(f),(N)}$: unless $t_w^{(f),(N)}$ is
larger than the time $r$ needs to grow up to its fixed point $r^{low}$, the effective decay will be longer than $t_r^{(N)} +t_{o}^{\textrm{final}}$. 
 We can estimate at which $t_r$ we expect the $n(t=t_r) = w^{*}_{(iii)} e^{-\delta \tau} + r^{*}_{(iii)} e^{-\delta \tau}  e^{\lambda_r t^{\textrm{off}}}$ to be lower than $n^{*}_{(iii)}$: this occurs for $t_{w}^{(f)} < (1/\lambda_r) \ln(n^{*}_{(iii)}/m^{*}_{(iii)}) + \tau \delta/\lambda_r$, which corresponds to $t^{\textrm{off}}<16$ for the sequence with two pulses where $2\tau^{(N=2)}=\tau=60$, and thus $2t_{r}^{(N=2)}>32$. Indeed, Fig. 1 shows that for $t_r>40$ the sequence with two pulses performs better than the single pulse, as here the second pulse of the two pulse sequence is lower. 
 
 For this $t_r$, a small increase in $s$ from $s=-1$ implies the presence of the $t_w^{(i),(N)}$, which means that $r$ regrows faster during the $t_w^{(f),(N)}$ phase, as well as a shorter $t_w^{(f),(N)}$. Thus, the first pulse in the sequence will yield a lower minimum again, and then certainly the single pulse is better.

\subsection*{5. Phase space trajectories (Fig.~5 a) for skewnesses $s=-0.5$, $s=0.2$ and $s=0.5$}
\begin{figure}[H]
  {\includegraphics[width = 0.95\linewidth]{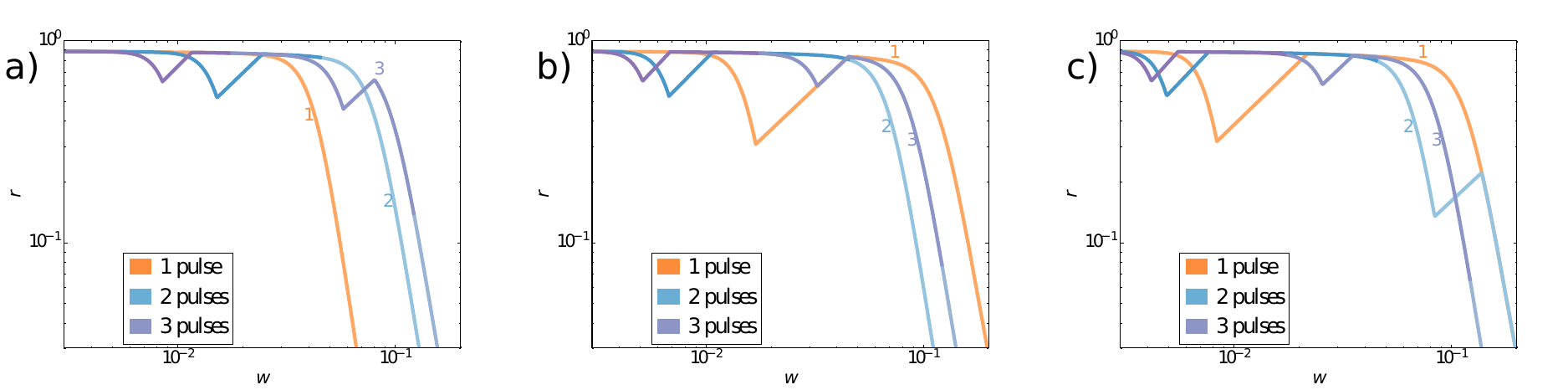}}
	\caption{\label{fig:trajectories_different_skewness} 
Phase space trajectories for pulse sequences with $\tau=60$ and $t_r=10$, consisting of one, two or three pulses, respectively. Three different skewnesses are considered: $s=-0.5$ (a), $s=0.2$ (b) and $s=0.5$ (c). As in Fig.~5 a the trajectories start at the \textit{(free)} fixed point close to the $w$-axis (beyond the limits of the figure) and $r/w$ steadily increases from pulse to pulse. Furthermore, the population composition at the end of the pulse sequences is similar for one, two or three pulses. 
	}
\end{figure}

In Fig.~5 a we show phase space trajectories for pulse sequences with $\tau=60$, $t_r=10$ and skewness $s=0.9$, consisting of one, two or three pulses, respectively. The trajectories start at the \textit{(free)} fixed point close to the $w$-axis and evolve towards the $r$-axis. $r/w$ increases from pulse to pulse since the more resistant species $r$ steadily takes over the population. Furthermore, the population composition at the end of the pulse sequence is similar for all three trajectories. The same holds true for other skewnesses. In Fig.~\ref{fig:trajectories_different_skewness} we show exemplary phase space trajectories for skewnesses $s=-0.5$, $s=0.2$ and $s=0.5$.

\subsection*{6. Constant amount of antibiotics: Estimate}

\begin{figure}[H]
  {\includegraphics[width = 0.95\linewidth]{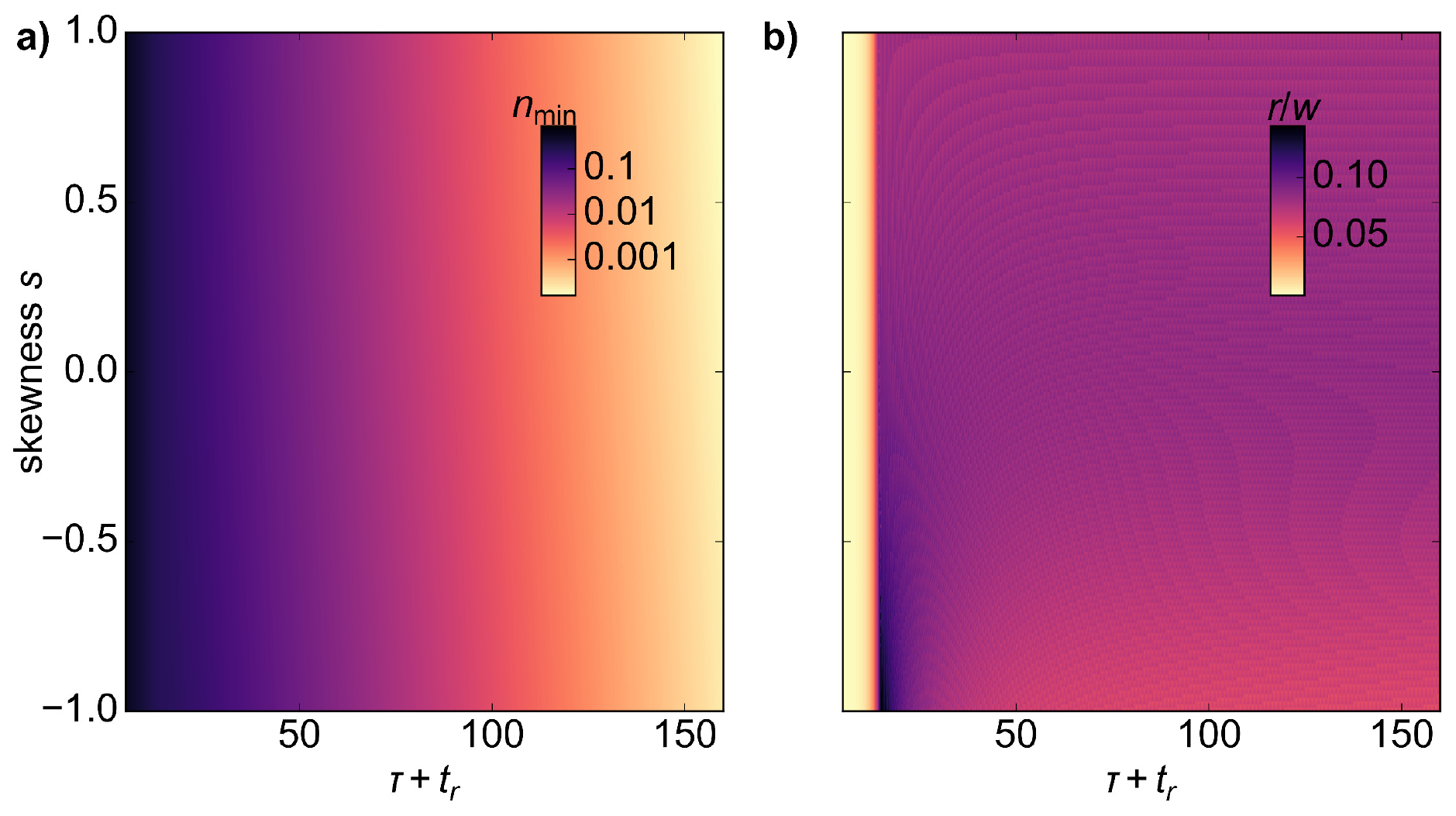}}
	\caption{\label{fig:constsum} 
Population minimum $n_{\mathrm{min}}$ (a) and composition $r/w$ (b) for a single pulse of fixed $\tau + t_r$ instead of constant $\tau$ and fixed $t_r$. Both quantities show little dependency on the skewness $s$. For this case, there are two competing effects: the smaller $t_r$, the less the resistant species is depleted, but simultaneously the smaller $t_r$ the larger $\tau$ and thus, the wild-type is depleted longer. As a result, not only optimising the onset but also avoiding a large offset is beneficial.
	}
\end{figure}

As shown in Fig.~\ref{fig:constsum}, keeping $t_\mathrm{sum}=\tau+t_r$ constant
shows that the dependency on the skewness is much less
pronounced than for keeping $\tau$ constant, which we explain in the following.

For the case of fixed $t_\mathrm{sum}$, there are two competing effects: on the one hand, for small $t_r$, $\tau=t_\mathrm{sum}-t_r$ is larger than for large $t_r$, and thus for small $t_r$ the population $w$ can be depleted longer. 
On the other hand, larger $t_r$ increase the chance to reduce the resistant species $r$, and thus also $n$. 
Thus, for $s=1$, where the pulse starts with the \textit{(low)} regime (see Fig. 2 in the main text),
it makes sense to exploit the full optimal onset time, such that $\tau= t_r + t_o$, since then $r$ has not grown much until the
beginning of the \textit{(high)} regime. This then means that $t_{\mathrm{sum}} = 2\tau-t_o$, and thus $\tau \approx t_{\mathrm{sum}}/2+7.7$. 
On the contrary, for $s=-1$ when the pulse starts with the \textit{(high)} regime, it is best to use
$t_w^{(f)}\approx t_o^{\mathrm{final}}(t_w^{(i)}=0, \mathrm{large}\ t_r)$ and so $t_r\approx \tau-t_o^{\mathrm{final}}(t_w^{(i)}=0, \mathrm{large}\ t_r)$ in order to prevent the resistant species from taking over the population in the subsequent \textit{(low)} regime. So, it is ideal to use $\tau\approx t_{\mathrm{sum}}/2+ t_o^{\mathrm{final}}(t_w^{(i)}=0, \mathrm{large}\ t_r)/2\approx t_{\mathrm{sum}}/2 $.
Together, these two strategies mean the pulses are optimised in such a way that their durations do not differ drastically
for the different skewnesses. Thus, the best minimum and the mutant/wild-type ratio is approximately the same for all 
skewnesses with a little improvement (deterioration) towards skewness s = 1 for the minimum (ratio).
This is also reflected in the ratio $r/w$ in Fig\ref{fig:constsum}b, which sharply increases until it makes sense to use $t_r>0$, namely
at $\tau + t_r=t_o$. For low $s$, the ratio $r/w$ is slightly higher, as the \textit{(low)} regime where the resistant species can grow, is at the end
of the pulse. For higher values of $\tau + t_r$, the ratio is very similar across all skewnesses.

\end{document}